\documentclass[aps,pra,twocolumn,superscriptaddress,notitlepage]{revtex4-2} 

\usepackage{graphicx}
\usepackage{mathptmx, textcomp}
\usepackage[usenames]{xcolor}
\usepackage{dcolumn}
\usepackage{bm}
\usepackage{amssymb, amsmath}
\usepackage{physics}
\usepackage{verbatim}
\usepackage{amsmath}
\usepackage[normalem]{ulem}
\usepackage{lineno}
\usepackage{float}
\usepackage{comment}
\usepackage[utf8]{inputenc}
\usepackage[colorlinks,urlcolor=blue,citecolor=blue,linkcolor=blue]{hyperref}
\usepackage{cleveref}

\newcommand{\impfrac}{\chi}

\begin{document}

\title{Three-body Physics in the Impurity Limit of $^{39}$K Bose-Einstein Condensates}

\author{A. \ M. \ Morgen}
\affiliation{Center for Complex Quantum Systems, Department of Physics and Astronomy, Aarhus University, Ny Munkegade 120, DK-8000 Aarhus C, Denmark}
\author{S.\ S.\ Balling}
\affiliation{Center for Complex Quantum Systems, Department of Physics and Astronomy, Aarhus University, Ny Munkegade 120, DK-8000 Aarhus C, Denmark}
\author{M.\ T.\ Strøe}
\affiliation{Center for Complex Quantum Systems, Department of Physics and Astronomy, Aarhus University, Ny Munkegade 120, DK-8000 Aarhus C, Denmark}
\author{T.\ G.\ Skov}
\affiliation{Center for Complex Quantum Systems, Department of Physics and Astronomy, Aarhus University, Ny Munkegade 120, DK-8000 Aarhus C, Denmark}
\author{M.\ R.\ Skou}
\affiliation{Center for Complex Quantum Systems, Department of Physics and Astronomy, Aarhus University, Ny Munkegade 120, DK-8000 Aarhus C, Denmark}
\author{A.\ G.\ Volosniev}
\affiliation{Center for Complex Quantum Systems, Department of Physics and Astronomy, Aarhus University, Ny Munkegade 120, DK-8000 Aarhus C, Denmark}
\author{J. J. Arlt}
\affiliation{Center for Complex Quantum Systems, Department of Physics and Astronomy, Aarhus University, Ny Munkegade 120, DK-8000 Aarhus C, Denmark}
\date{\today}

\begin{abstract}

Loss spectroscopy is a key tool for investigating systems where important system parameters are linked to intrinsic resonant loss processes. We investigate loss processes of impurity atoms embedded in a medium of a Bose-Einstein Condensate close to a Feshbach resonance. In this case, three-body loss processes occur faster than the measurement duration, impeding a direct time-resolved measurement. Here, we discuss how an even faster two-body loss process can be used to probe the system. The time-dependent number of atoms in the medium is reconstructed from such measurements, allowing for the extraction of the three-body loss rate coefficient $L_3$ and its scaling with scattering length. Moreover, the medium atom number is reconstructed from spectroscopic loss measurements. This allows for a comparison of the medium densities based on both the extracted loss rates and the spectroscopically reconstructed atom number. Finally, the number of lost medium atoms per loss event is evaluated and found to exceed 2 at strong interactions, which is attributed to secondary collisions in the medium. These investigations establish the use of a fast loss mechanism as a new tool in the field and provide quantitative measurements of three-body losses at large interaction strengths.

\end{abstract}

\maketitle

\section{Introduction}

Ultracold atomic gases offer a unique platform for investigating few-body physics in a controlled environment. Typically, they feature a combination of elastic and inelastic collisions, which depend on the chosen states in single or multi-component ensembles. In many systems the inelastic loss rates can be resonantly enhanced by tuning the experimental parameters, such as the magnetic field or frequency of an external probe field. In fact, such resonances in the observed atom loss are an important signature when investigating Feshbach resonances~\cite{Chin2010Feshbach,Gerken2019}, Efimov physics~\cite{kraemer2006,zaccanti2009}, cold molecules~\cite{Wu2011,Wang2007,Klempt2008}, and Rydberg states~\cite{Duverger2024}. 

In Bose-Einstein Condensates (BECs), inelastic collisional processes are usually enhanced due to the high atomic density. Consequently, the time scale of the loss processes is relatively short, complicating the quantitative, time-resolved investigation of losses. In particular,  when the interactions within a BEC are tuned near a Feshbach resonance, three-body recombination processes can occur faster than the time required to observe the atoms. Thus, only the final number of atoms remaining after all losses have occurred can be recorded in such experiments. 

This is typically the case for the investigation of impurity physics with BECs in the vicinity of a Feshbach resonance between two quantum states. In this scenario, atoms in one state play the role of a strongly interacting impurity, which allows for the description of the system in terms of a polaron quasiparticle. Losses in these systems have been exploited in spectroscopic and interferometric measurements to infer properties of the impurity~\cite{jorgensen2016, ardila2019, Skou2020, Morgen2023, etrych2024}. However, the details of the three-body recombination losses have not been analysed in detail within this regime so far. 

Note that previous work on loss processes of impurities has used single atoms in $^{87}\text{Rb}-^{133}\text{Cs}$~\cite{Spethmann2012} and $^{41}\text{K}-^{87}\text{Rb}$~\cite{Hewitt2024}. The impurity limit was used to single out a specific loss channel, which is also the case in our system, which features strongly interacting impurities in a dense BEC of $^{39}\text{K}$ atoms.

\begin{figure}[t!]
    \centering
    \includegraphics[width=\columnwidth]{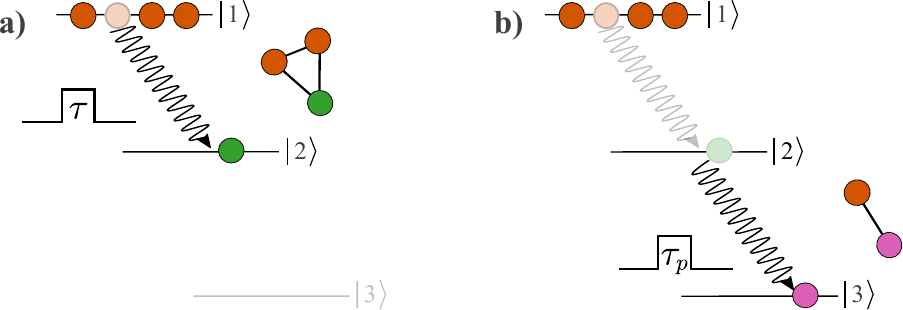}
    \caption{Creation and loss measurement in a strongly interacting many-body state. \textbf{a)} The first rf pulse, $\tau$, transfers an admixture of atoms from the BEC state, $\ket{1}$, to the strongly interacting impurity state $\ket{2}$. The most relevant loss between these two states occurs due to three-body recombination between two BEC atoms and one impurity atom. \textbf{b)} The second rf pulse, $\tau_p$, transfers all remaining impurity atoms to the third state, $\ket{3}$. The dominant loss channel between $\ket{1}$ and $\ket{3}$ are two-body spin-exchange collisions.}
    \label{fig:injection_ejection}
\end{figure}

A quantitative investigation of the loss rate due to three-body recombination of impurity and medium atoms in a BEC requires a time-resolved measurement of at least one component's atom number.
Here, we show that this is possible for the majority BEC component by using a third state in addition to the impurity and BEC states in a so-called ejection sequence~\cite{Yan2020}. This sequence is illustrated in Fig.~\ref{fig:injection_ejection}. It consists of two radio frequency (rf) pulses. The first pulse transfers an admixture from the medium BEC state, $\ket{1}$, to the impurity state, $\ket{2}$, thus creating a strongly interacting mixture. Within the following evolution time, impurity and medium atoms interact at the chosen interaction strength and both impurity and medium atoms are lost through three-body recombination. The interacting mixture is probed by a second, short rf pulse which transfers all remaining impurity atoms to a third state, $\ket{3}$. In an ideal case, the third state would be non-interacting with the medium state, such that the remaining atoms in both states can be measured. However, this is not the case for our system. Instead, the remaining impurities in the third state are lost through two-body spin-exchange collisions with medium atoms in state $\ket{1}$. Nonetheless, the difference in the two loss processes allows for a quantitative, time-resolved measurement of the three-body recombination rate. 

In this paper, we show that the ejection sequence above and the subsequent fast two-body loss allow for a time-resolved measurement of the medium atom number. This is utilized for analysing the three-body recombination rate of an impurity and two BEC atoms at strong interactions. Thus, it provides insight into the few-body physics of the system and allows for accurate modelling of the losses. This makes it possible to account for the decay in the BEC atom number and the corresponding atomic density of the system. The atomic density determines the scaling parameters in the system, which are crucial for interpreting experimental results. In addition, it is shown that the medium atom number can be reconstructed from spectroscopic measurements, and the obtained medium densities based on both methods are compared. Importantly, the number of lost medium atoms per loss event can be evaluated, showing the presence of significant secondary collisions at large interaction strengths.

The paper is organized as follows. After this introduction,  an overview of the experimental system is provided in Sec.~\ref{sec:experiment}. Section~\ref{sec:reconstruction} describes the reconstruction of the medium atom number based on the observed signal in the experimental sequence. In Sec.~\ref{sec:density}, the method for calculating the medium atom number is used to extract the atomic density of the system. The three-body loss rate is analysed and compared to theoretical predictions in Sec.~\ref{sec:thee-body}. Section~\ref{sec:loss_meachnisms} provides a detailed analysis of the number of medium atoms lost per impurity. Finally, a conclusion is provided in Sec.~\ref{sec:conclusion}.

\section{Experimental implementation} 
\label{sec:experiment}

Our experiment is performed with BECs of $^{39}$K, produced in an optical dipole trap (ODT)~\cite{wacker2015}. In brief, a dual-species magneto-optical trap captures and cools $^{39}$K and $^{87}$Rb simultaneously. Subsequently, optical molasses and pumping are applied, and both species are magnetically trapped in the $\ket{F=2,m_F=2}$ state, where $F$ and $m_F$ are the total angular momentum quantum number and its projection, respectively. Microwave radiation is then used to selectively evaporate $^{87}$Rb atoms, which cools $^{39}$K atoms sympathetically. When all $^{87}$Rb atoms are evaporated, the remaining $^{39}$K atoms are loaded into a crossed-beam optical dipole trap. At this stage, rapid adiabatic passages are performed to transfer the $^{39}$K atoms to the $\ket{F=2,m_F=-2}$ state, and further to the $\ket{F=1,m_F=-1}$ state. The final evaporation towards condensation is performed in the ODT by lowering the power of the two beams at a magnetic field of approximately 41G, where the rethermalization is enhanced due to the presence of the Feshbach resonance at 33.64~G~\cite{roy2013}.

In the following, medium atoms are in the $\ket{F=1,m_F=-1} \equiv \ket{1}$ state, while the impurity state is $\ket{F=1,m_F=0} \equiv \ket{2}$, and the third state, $\ket{F=1,m_F=+1}\equiv \ket{3}$, is used in the ejection sequence, as illustrated in Fig.~\ref{fig:injection_ejection}. An interstate Feshbach resonance at $113.8~$G, between the medium and impurity state~\cite{lysebo2010,Tanzi2018}, allows for tuning their interaction, characterized by the impurity-medium scattering length, $a$. Close to this resonance, both the medium-medium scattering length, $a_B = 10a_0$, and the impurity-impurity scattering length, $a_I = -25a_0$, are small~\cite{lysebo2010,Tanzi2018}.

The initial rf pulse has a duration of $\tau \approx$1~µs and is applied at the resonance frequency between the $\ket{1}$ and $\ket{2}$ states, creating a coherent 10\% admixture of atoms in the impurity $\ket{2}$ state~\footnote{The resonance frequency and the necessary pulse duration and power are determined independently using thermal atoms.}. The following evolution time $t_e$ has a variable duration, allowing for interaction between the impurity admixture and the medium. It is interrupted when the system is probed by the second rf $\pi$-pulse, which transfers the remaining impurity atoms to the $\ket{3}$ state. 

\begin{figure}[t!]
    \centering
    \includegraphics[width=\columnwidth]{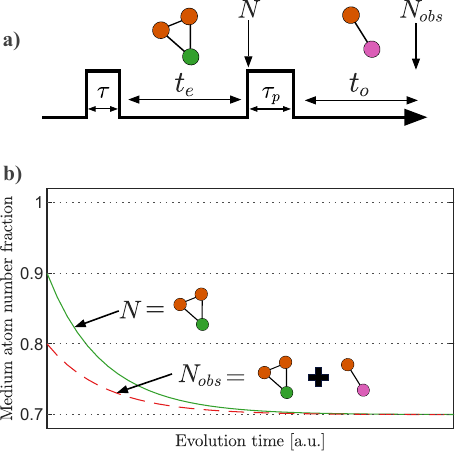}
    \caption{\textbf{a)} Overview of the measurement sequence. The two rf pulses are separated by the evolution time, $t_e$, followed by a time $t_o$ until the BEC atom number is observed. During the evolution time, three-body recombination takes place and reduces the number $N$ of BEC atoms. After the second rf probe pulse, two-body collisions take place and further reduce the number of BEC atoms to  $N_{obs}$. \textbf{b)} Illustration of the real $N(t)/N_0$ (green solid line) and observed $N_{obs}(t)/N_0$ (red dashed line) medium atom number fractions as a function of the evolution time. The initial atom number is $N_0$ and the assumed impurity fraction is $\chi = 10\%$, leading to the offset value.  The green line is the result of pure three-body recombination between medium and impurity atoms. The red dashed line contains the additional loss due to two-body collisions and corresponds to the experimentally observed atom number. 
    }
    \label{fig:loss_measurement}
\end{figure}

After the ejection sequence, the atoms are briefly held in the ODT to ensure that all two-body losses have taken place. Subsequently, the optical potential is turned off to start a time-of-flight expansion with a total duration of 28~ms. Finally, the number of remaining atoms in the medium state is recorded using absorption imaging~\cite{Ketterle1999,Reinaudi2007,Vibel2024}. First, repumping light is applied to transfer the atoms into the $\ket{F=2}$ manifold. Then, detection light resonant with the $\ket{F=2}$ to the excited $\ket{F'=3}$ transition is used to record an absorption image on a charged-coupled device camera, which allows for the extraction of the atom number and peak density.  

To characterize the interaction strength in ultracold quantum gases, the following parameters associated with the atomic density, are typically used. The characteristic energy is given by $E_n = \hslash^2 k_n^2/2m$ with wavenumber, $k_n = (6\pi^2 n)^{1/3}$, where $n$ is the peak atomic density of the medium. This also leads to the impurity-medium inverse interaction strength typically characterized by $1/k_n a$. In the experimental sequence described above, the relevant medium atom number, $N$, for calculating the atomic density is the one at the time of applying the second rf pulse. However, this atom number is not the one observed, $N_{obs}$, at the end of the experimental sequence, as illustrated in Fig.~\ref{fig:loss_measurement}. Instead, the observed atom number is lower, since all impurities are lost before detection, due to the two- or three-body loss processes discussed in the previous section. In order to extract the relevant medium atom number, it is thus necessary to reconstruct the medium atom number at a given evolution time $t_e$ in the experimental sequence based on the observed atom number.

\section{Atom number reconstruction} 
\label{sec:reconstruction}

In the following, a method for calculating the medium atom number from the observed number is presented, which we term atom number reconstruction. This method is first applied in time-resolved loss measurements, providing a medium atom number after a variable evolution time. Secondly, it is applied in frequency resolved spectroscopic measurements to infer the medium atom number at the point of probing the system.

\subsection{Decay measurements and reconstruction}
\label{sec:decay}

To calculate the medium atom number after a variable evolution time $t_e$, the loss rate associated with the interaction is required. Once this rate is known, the medium atom number at any time $t_e$ can be determined from the initial atom number. 

The number of medium atoms is measured using the ejection sequence outlined in Fig.~\ref{fig:injection_ejection}. The expected outcome of such a measurement is illustrated in Fig.~\ref{fig:loss_measurement}, showing the real and an observed atom number. The real atom number  $N(t)$ decays primarily due to three-body recombination between atoms in the $\ket{1}$ and $\ket{2}$ states. The observed atom number $N_{obs}(t)$ is lower due to the additional two-body losses between atoms in the $\ket{1}$ and $\ket{3}$ states after the second rf probe pulse.

The decay process can be modelled as follows. For a BEC with $N_0$ atoms initially, the first rf pulse is adjusted to populate the impurity state with a fraction $\impfrac$. Thus, the number of the medium atoms starts at $N_0(1-\impfrac)$ and decreases to $N_0(1-3\impfrac)$ for long evolution times, assuming a factor 3 due to three-body losses. The observed medium atom number has additional losses due to the transfer of the impurities to the $\ket{3}$ state in the ejection sequence. In particular, for short evolution times, the observed atom number starts at $N_0(1-2\impfrac)$ where the factor 2 reflects the additional two-body loss. However, for long evolution times, all impurity atoms decay due to three-body loss, and the ejection sequence has no effect. Thus, the observed atom number also decays to $N_0(1-3\impfrac)$ for long evolution times. Based on this understanding, it is possible to reconstruct the real medium atom number from the observed one. 
This reconstructed number then only reflects the loss process between the medium and impurity state, corresponding to the real decay mechanism, which is of interest here.

The number of medium atoms at a variable time is given by
\begin{equation}
    N(t) = N_0(1-\impfrac) - \eta_3(N_0\impfrac - N_i(t)),
    \label{eq:Atom_number}
\end{equation}
where $\eta_3$ is the number of medium atoms lost per impurity, $N_i(t)$ is the remaining number of impurity atoms at time $t$, $N_0$ is the initial BEC atom number and $\impfrac$ is the initial impurity fraction. The first term corresponds to the offset mentioned above, while the second term is given by the number of impurity atoms lost at time t, $N_0\impfrac - N_i(t)$, times the loss coefficient $\eta_3$. Here, $\eta_3$ is expected to be $\approx 2$ for three-body recombination.
The probe pulse transfers the remaining impurities to the third state, $\ket{3}$, where additional two-body losses decrease the number of observed medium atoms to
\begin{equation}
    N_\text{obs}(t) = N(t) - \eta_2 N_i(t),
    \label{eq:N_obs}
\end{equation}
where $\eta_2$ is the number of medium atoms lost per impurity after their transfer to the $\ket{3}$ state, expected to be $\approx 1$ for two-body losses. Rearranging Eq.~\eqref{eq:N_obs} allows the number of impurities at time $t$ to be expressed in terms of $N_\text{obs}(t)$ and $N(t)$, as $N_i(t) = (N(t)-N_\text{obs})/\eta_2$. For long evolution times, all impurities are lost through three-body recombination before the probe pulse is applied and the observed atom number is $N_\infty \equiv N(t\rightarrow \infty) = N_0(1-\impfrac) - \eta_3N_0\impfrac$. Inserting these expressions for $N_i(t)$ and $N_\infty$, into Eq.~\eqref{eq:Atom_number} gives,
\begin{figure}[t!]
    \centering
    \includegraphics[width=\columnwidth]{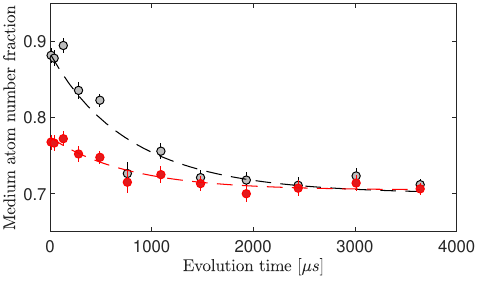}
    \caption{Observed, $N_\text{obs}(t)/N_0$, (red circles) and reconstructed, $N(t)/N_0$, (gray circles) medium atom number fractions at $a=-200a_0$, fitted with the exponential decay of Eq.~\eqref{eq:old_density_obs} (red dashed line) and the differential equation in Eq.~\eqref{eq:Diff_final} (gray dashed line), respectively. From the latter, the three-body recombination rate coefficient, $L_3$, is extracted.}
    \label{fig:decay_curve}
\end{figure}

\begin{equation}
    N(t) = \frac{1}{\eta_3-\eta_2}\left(\eta_3N_\text{obs}(t) - \eta_2N_\infty\right).
    \label{eq:N_final}
\end{equation}
Thus, the medium atom number $N(t)$ can be reconstructed from the observed medium atom number, $N_\text{obs}(t)$, based on the observed medium atom number at long evolution times, $N_\infty$. Moreover, the coefficients $\eta_2$ and $\eta_3$ are obtained experimentally from the observed medium atom number. This is possible, since for short evolution times only two-body losses are relevant, while for long evolution times only the effect of three-body losses is observed. 

An example of the observed medium atom number and the corresponding reconstructed medium atom number is shown in Fig.~\ref{fig:decay_curve} at low impurity-medium scattering length. As expected, the reconstructed medium atom number starts at $0.9N_0$ and decreases to $0.7N_0$ at long times~\footnote{Additionally, we can also extract the temperature of the small thermal part of our cloud, for which we observe a relative increase in temperature of $\sim 0.1~\text{T}/\text{T}_\text{C}$, from short to long evolution times.}. 

The reconstructed medium atom number from Eq.~\eqref{eq:N_final} allows us to estimate the three-body recombination loss rate. The starting point  for this analysis is the differential equation for the number of medium atoms
\begin{align}
    \begin{split}
            \frac{dN_B}{dt} = &-\frac{2}{3}L_3^{BBI}\int n_B^2n_I d^3r - \frac{1}{3}L_3^{BII}\int n_B n_I^2 d^3r \\
            &- L_3^{BBB}\int n_B^3 d^3r - N_B/\tau,
    \end{split}
    \label{eq:Diff_full}
\end{align}
where the indices $B$ and $I$ refer to the medium and impurity, with atomic densities $n_B$ and $n_I$~\cite{wacker2016,Mikkelsen2015}. The equation includes three loss channels with the loss rate coefficients $L_3$ specified for each particular loss channel. Losses due to collisions with background atoms are modelled by the last term, $-N_B/\tau$. In principle, a similar differential equation applies for the number of impurity atoms, leading to coupled equations. However, the above differential equation is simplified by the following considerations. The density of the impurities is much smaller than the density of the BEC, which means that the second term in Eq.~\eqref{eq:Diff_full} may be neglected. Additionally, the scattering length between the BEC atoms is very small, $a_B = 10a_0$, and the timescales associated with the measurements are much shorter than those associated with background collisions. Therefore, we can also neglect the third and fourth terms in Eq.~\eqref{eq:Diff_full}. 

These considerations simplify Eq.~\eqref{eq:Diff_full} to
\begin{equation}
    \frac{dN}{dt} = -\frac{2}{3}L_3\int n_B^2n_I d^3r,
    \label{eq:Diff_short}
\end{equation}
with $L_3 \equiv L_3^{BBI}$ and  $N\equiv N_B $. Since the first rf pulse generates an impurity admixture homogeneously within the BEC, the density of the impurities at $t=0$ is determined by the density of the BEC~\cite{ardila2019}. This motivates the choice $n_I(r,t) = \rho (t) n_B(r,t)$ where $\rho(t)$ is a proportionality function, which is initially $\rho(0)\approx\chi$ (for small $\chi$) and tends to $\rho(\infty)=0$. 

This drastically simplifies the analysis, and assumes that the spatial impurity distribution retains the BEC shape, corresponding to a single-mode approximation. Based on this assumption,  Eq.~\eqref{eq:Diff_short} yields
\begin{equation}
    \frac{dN}{dt} = -\frac{2}{3}L_3\rho(t) \int n_B^3d^3r.
    \label{eq:Diff_short2}
\end{equation}
Using the Thomas-Fermi approximation for the density distribution of the BEC, the above integral has a well-known solution~\cite{soding1999}. The density fraction of impurity atoms in the system can be approximated in terms of the medium and impurity atom numbers as
\begin{equation}
    \rho(t) = \frac{N_i(t)}{N(t)},
    \label{eq:c_fraction1}
\end{equation}
where $N_i(t)$ and $N(t)$ are the number of impurity and medium atoms at time $t$, respectively. The relation between these is given by Eq.~\eqref{eq:Atom_number}, and inserting $N_i(t)$ into Eq.~\eqref{eq:c_fraction1} yields
\begin{equation}
    \rho(t) = \frac{N(t)/N_0 +\impfrac(1+\eta_3) -1}{\eta_3N(t)/N_0}.
    \label{eq:c_fraction}
\end{equation}
With these results, Eq.~\eqref{eq:Diff_short2} gives the following differential equation for the decay of the medium atom number 

\begin{equation}
    \begin{split}
        \frac{dN}{dt} = -\frac{2}{3}&L_3\frac{7}{6}\frac{15^{4/5}}{(14\pi)^2}\left(\frac{m\overline{\omega}}{\hslash}\right)^{12/5}\frac{1}{a_B^{6/5}}N^{9/5} \\
        &\times \frac{N/N_0 +\impfrac(1+\eta_3) -1}{\eta_3N/N_0},
    \end{split}
    \label{eq:Diff_final}
\end{equation}
where $a_B$ is the medium-medium scattering length and ${\overline{\omega} = \left(\omega_x\omega_y\omega_z\right)^{1/3}} = 2\pi\times75.2(1)$ Hz is the average trap frequency.

Integrating the differential equation~\eqref{eq:Diff_final} yields a transcendental equation that relates $N$ and $t$. We noticed, however, that a direct numerical integration of Eq.~(\ref{eq:Diff_final}) leads to a solution at almost the same computational cost. Therefore, Eq.~(\ref{eq:Diff_final}) is solved numerically and its solution is fitted to the reconstructed medium atom number, as shown in Fig.~\ref{fig:decay_curve}, with $L_3$ as a free parameter. The three-body loss rate measurement is repeated at various attractive impurity-medium scattering lengths to map out the dependency of $L_3$ on $a$. The extracted three-body loss rate coefficients are discussed further in Sec.~\ref{sec:thee-body}.

\subsection{Spectroscopic measurements and reconstruction}
\label{sec:spectroscopy}

\begin{figure}[t!]
    \centering
    \includegraphics[width=\columnwidth]{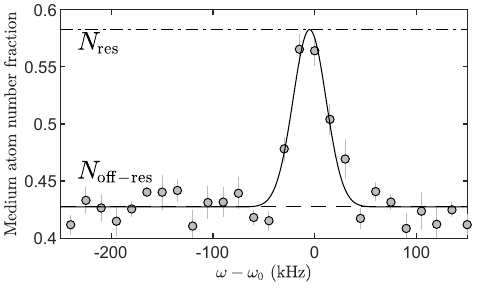}
    \caption{Observed medium atom number fraction as a function of the frequency of the ejection probe pulse, $\omega$, relative to the resonance frequency between the $\ket{2}$ and $\ket{3}$ states, $\omega_0$. This spectrum was recorded at an impurity-medium scattering length of $a = -200~a_0$, with impurity fraction $\chi \approx 20\%$, and an evolution time of $t_e = 20$~µs. The observed loss of medium atom numbers is fitted with a Gaussian (solid line). Far off-resonance, the observed loss is due to three-body recombination (dashed line). Close to resonance, the observed loss is a mixture of two-body and three-body losses (dash-dotted line).}
    \label{fig:spectroscopy}
\end{figure}

In the experimental sequence discussed above, the evolution time between the two rf pulses was varied, while their frequencies were fixed. An alternative approach is to vary the frequency of the second, ejection rf pulse, $\tau_p$, to obtain the spectral response of the impurity atoms for a fixed evolution time $t_e$. An example of such an ejection spectrum is shown in Fig.~\ref{fig:spectroscopy}. 
In this case, the ejection rf probe pulse duration is $\tau_p = 20~$µs to obtain appropriate frequency resolution~\footnote{It remains a $\pi$-pulse by lowering the power appropriately.} while keeping the duration shorter than the expected lifetime of the impurities~\cite{SkouPRA2022}. An impurity fraction of $\chi \approx 20\%$ was chosen to obtain a sufficient spectroscopic signal.

For such spectroscopic datasets, calculating the medium atom number at the point of probing the system is also possible, as explained in the following. If the probe pulse is off-resonant, all impurities remain in the $\ket{2}$ state and are lost through three-body recombination, and the observed atom number off-resonance is therefore, $N_{\text{off-res}} = N_\infty$. If the probe pulse is on resonance, the remaining impurities are transferred to the $\ket{3}$ state after the evolution time, and the observed atom number corresponds to $N_{\text{res}} = N_\text{obs}$. The observed atom numbers off- and on-resonance can now be inserted into Eq.~\eqref{eq:N_final}, which then takes the form, 
\begin{equation}
    N(t) = N_{\text{off-res}} + \frac{\eta_3}{\eta_3-\eta_2}\left(N_\text{res} - N_{\text{off-res}}\right).
    \label{eq:true_atomnumber_ejec_0}
\end{equation}
This form has the advantage that the loss coefficient, $\eta_2$, only appears in the denominator of the second term. Typically, $\eta_2$ cannot be extracted for a single ejection spectrum and instead, we assume that $\eta_3 - \eta_2 = 1$. This is true under the assumption that these are strictly three- and two-body loss coefficients. Based on this approximation, the medium atom number at the point of probing the system is
\begin{equation}
    N(t) = N_{\text{off-res}} + \eta_3\left(N_\text{res} - N_{\text{off-res}}\right).
    \label{eq:true_atomnumber_ejec}
\end{equation}
Thus, the medium atom number $N(t)$ can be reconstructed from spectroscopic observation of $N_\text{res}(t)$, and $N_\text{off-res}(t)$. The coefficient $\eta_3$ can also be obtained experimentally, since the atom number observed off-resonance contains information only about the three-body losses. Further details of $\eta_2$ and $\eta_3$ are provided in Sec.~\ref{sec:loss_meachnisms}.

Equations~\eqref{eq:Diff_final} and~\eqref{eq:true_atomnumber_ejec} provide a framework for reconstructing the medium atom number. Importantly, the measurements in Sec.~\ref{sec:decay} allow for a full reconstruction of the three-body recombination loss process between medium and impurity atoms. The result of Eq.~\eqref{eq:true_atomnumber_ejec} is used in the context of spectroscopic measurements, allowing for a determination of the medium atom number at the time of probing the system, which is necessary for calculating the characteristic scaling parameters, $E_n$ and $k_n$.

\subsection{Empirical approach}
\label{sec:empirical}

Previously, an empirical solution was used to capture the decay of the medium atom atom number~\cite{Skou2020,Morgen2023}, which is included here briefly for completeness and comparison. In this approach, the number of medium atoms was approximated with an exponential decay, and the observed number was fitted with
\begin{equation}
    N_\text{obs}(t) = \alpha e^{-\Gamma t}+\beta,
    \label{eq:old_density_obs}
\end{equation}
where $\Gamma$ is the associated loss rate, which together with $\alpha$ and $\beta$, act as free fitting parameters. In Fig.~\ref{fig:decay_curve}, the fit of Eq.~\eqref{eq:old_density_obs} to the observed number of medium atoms is shown. The exponential fit captures the data well, and the extracted loss rate is used to obtain the medium atom number at all times, 
\begin{equation}
    N(t) = N_0(1-\chi)e^{-\Gamma t}.
    \label{eq:old_density}
\end{equation}
In Ref.~\cite{Skou2020}, the dependence of the loss rate, $\Gamma$, on the impurity-medium interaction strength was mapped out. It can be extracted for any value of $1/k_na$, and applied in Eq.~\eqref{eq:old_density}. However, the exponential decay above assumes a homogenous density distribution of the medium~\cite{Spethmann2012}, which is not the case for our system. Thus, the loss rate, $\Gamma$, only provides an overall phenomenological loss rate of the system. Furthermore, this loss rate includes both three-body and two-body effects since it is obtained from the observed medium atom number. Typically, this leads to an overestimate of the losses between the medium and impurity state.

In the following section, the methods for calculating the medium atom number based on the loss rate coefficient (see Eq.~\eqref{eq:Diff_final}), the spectroscopic measurements (see Eq.~\eqref{eq:true_atomnumber_ejec}), as well as the empirical approach (see Eq.~\eqref{eq:old_density}) are compared directly.

\section{Atomic density determination} 
\label{sec:density}

The previous section introduced three methods for extracting the medium atom number after a variable evolution time. In this section, these methods are compared to each other with the aim of calculating atomic densities, which determine the characteristic parameters $E_n$ and $k_n$ introduced in Sec.~\ref{sec:experiment}. 

The spectroscopic measurements described in Sec.~\ref{sec:spectroscopy} offer the most direct method to determine the number of medium atoms, since it is extracted directly from the observed ejection spectrum. In particular, the method only requires the observed off- and on-resonance medium atom numbers together with the coefficient $\eta_3$, as shown by Eq.~\eqref{eq:true_atomnumber_ejec}. The coefficient $\eta_3$ does require the initial atom number and impurity fraction to be known (see Sec.~\ref{sec:loss_meachnisms}). However, Eq.~\eqref{eq:true_atomnumber_ejec} only provides the atom number at the precisely chosen evolution time and interaction strength of the spectroscopic measurement. 

\begin{figure}[t!]
    \centering
    \includegraphics[width=\columnwidth]{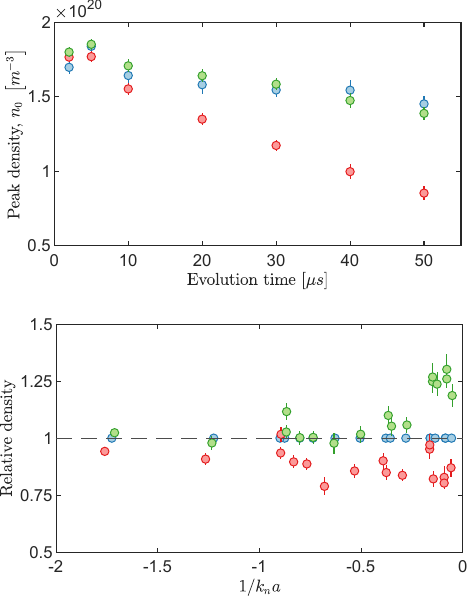}
    \caption{(top) Calculated peak densities for varying evolution time at $1/k_na = -0.35$ using spectroscopic measurements Eq.~\eqref{eq:true_atomnumber_ejec} (blue circles), the loss rate Eq.~\eqref{eq:Diff_final} (green circles), and an empirical method Eq.~\eqref{eq:old_density} (red circles). The densities are only given for the evolution times at which spectroscopic measurements were performed. (bottom) Calculated peak densities for varying interaction strengths at a fixed evolution time of $20$~µs. The result of Eq.~\eqref{eq:Diff_final} (green circles) and Eq.~\eqref{eq:old_density} (red circles) are shown relative to that of Eq.~\eqref{eq:true_atomnumber_ejec} (black dashed line).}
    \label{fig:densities}
\end{figure}

The decay measurements and the subsequent determination of the loss rate coefficient $L_3$ in Sec.~\ref{sec:decay}, provide a more general method to obtain the medium atom number. Based on the numerical solution of Eq.~\eqref{eq:Diff_final}, the atom number can in principle be obtained at any evolution time and any interaction strength. Besides the loss rate coefficient $L_3$, this calculation requires the initial atom number $N_0$, impurity fraction $\impfrac$, and lost atom number per impurity $\eta_3$, in addition to known experimental parameters. These parameters are obtained from independent measurements with BECs ($N_0$) and thermal atoms ($\impfrac$), and thus this method is less direct.

Finally, the empirical approach described in Sec.~\ref{sec:empirical} based on Eq.~\eqref{eq:old_density}, can be used to obtain the atom number. However, it does not include the limiting value of the medium atom number due to the small impurity fraction, and therefore, the atom number is typically underestimated at longer evolution times and strong interactions.

In the following, the results are given in terms of the peak density which determines $E_n$ and $k_n$. Assuming pure BECs with medium atom number, $N$, the peak density in the Thomas-Fermi approximation is given by
\begin{equation}
    n_{0} = \frac{15^{2/5}}{8\pi}\left(\frac{m\overline{\omega}}{\hslash \sqrt{a_B}}\right)^{6/5}N^{2/5},
\end{equation}
with the average trap frequency, $\overline{\omega}$, and the medium-medium scattering length $a_B$.
Figure~\ref{fig:densities} shows a comparison of the densities extracted for varying evolution times at a fixed, large impurity-medium interaction strength with $1/k_na = -0.35$.  

All extracted densities show the expected decay primarily due to three-body collision also discussed in Fig.~\ref{fig:decay_curve}. The three different calculations agree at short times. For the spectroscopic and the loss rate method, this agreement remains very good at all evolution times. In particular, both methods agree for long evolution times, where the loss is expected to slow down since the density continuously decreases. More surprisingly, they also agree at short evolution times, where the duration of the probe pulse is comparable to the evolution time. In this case, one may expect a longer effective evolution time is required in the loss rate method. Since both methods agree well at short evolution times, we conclude that effects due to finite pulse duration are small in the presence of fast two-body loss. In comparison, the empirical fit method clearly overestimates the losses at longer evolution times. Note, however, that the absolute differences in peak density remain moderate.

Figure~\ref{fig:densities} also shows a comparison of densities for varying the interaction strength at a fixed evolution time of $20$~µs. To highlight the differences in the three approaches, the results of the loss method Eq.~\eqref{eq:Diff_final} and the empirical method Eq.~\eqref{eq:old_density} are shown relative to the direct spectroscopic method Eq.~\eqref{eq:true_atomnumber_ejec}. For a wide range of weak interactions, the relative differences in these results are small. Close to unitarity, the results of the empirical method Eq.~\eqref{eq:old_density} underestimate the density in agreement with the top panel. Even closer to unitarity for $1/k_na < -0.2$, the calculated densities according to the loss method are higher than those reconstructed from spectroscopic measurements. This discrepancy indicates that additional loss processes play a role at very large scattering lengths, which is further discussed in Sec.~\ref{sec:loss_meachnisms}.

Importantly, the differences in atomic densities do not have a large effect on the inverse interaction strength $1/k_n a$, which is mainly determined by the chosen value of the scattering length $a$. The small differences in density only lead to the small horizontal offsets between data points as observed in Figure~\ref{fig:densities}~(bottom).

\section{Three-body loss analysis}
\label{sec:thee-body}
\begin{figure}[t!]
    \centering
    \includegraphics[width=\columnwidth]{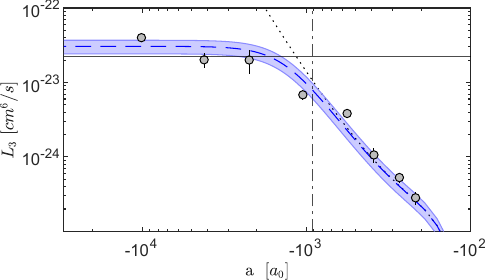}
    \caption{Extracted three-body loss rate coefficients (gray points) as a function of the impurity-medium scattering length, $a$. A fit to the data with Eq.~\eqref{eq:L_eff} (blue dashed line) captures the small scattering length behaviour given by Eq.~\eqref{eq:L_3} (black dotted line) and the saturation for large values of the scattering lengths. The error of the fit is indicated by the blue-shaded region. The inverse characteristic wavenumber of the medium, $1/k_n$ (vertical black dash-dotted line), indicates the scattering length at which the system enters a strongly interacting regime; for larger scattering lengths the loss rate is expected to reach its maximal value determined by the density of the medium. Additionally, the expected saturation value of the loss rate coefficient given by Eq.~\eqref{eq:L_max} is also shown (horizontal solid black line).}
    \label{fig:K3_coeffs}
\end{figure}

To understand the three-body physics at play, we investigate the three-body loss rate coefficient, $L_3$ (see Sec.~\ref{sec:decay}), in this section. The extracted loss rates are presented in Fig.~\ref{fig:K3_coeffs}, and are observed to increase as a function of $|a|$, eventually saturating in the unitarity limit.
The loss rate coefficient is expected to follow a behaviour given by (cf. Ref.~\cite{Helfrich2010})
\begin{equation}
    \frac{1}{2!}n_lC\frac{\hslash}{m}a^4\frac{\sinh(2\eta_*)}{\sin(s_0\ln(a/a_-))^2 + \sinh(\eta_*)^2}.
    \label{eq:L_3}
\end{equation}
This expression consists of two terms, the first one $n_lC\frac{\hslash}{m}a^4$ describes the universal $a^4$ behaviour where $n_l$ is the number of atoms in the collision, usually set to 3, and $C$ is a dimensionless constant, which acts as a free fitting parameter in our analysis (see Ref.~\cite{weber2003}). The second term incorporates Efimov physics, which has a characteristic log-periodic behaviour of the scattering length. For our experimental setting, the scaling is set by $s_0 = 0.413$~\cite{Helfrich2010}. The other parameters are not known for our system. An expectation for $\eta_*$ follows from previous studies of Efimov physics in $^{39}$K~\cite{roy2013,wacker2018}, setting it to $\eta_* \simeq 0.2$. The three-body parameter can be estimated from a two-channel model, $|a_-| = 3\times10^5 a_0$~\cite{jorgensen2016,Levinsen2015}.

Additionally, the use of a BEC is expected to lead to a reduction in the observed three-body loss rate, compared to the case of a thermal cloud~\cite{Burt1997}. In fact, for a thermal gas, the probability of finding three atoms close together is $3!$ times higher compared to an ideal BEC~\cite{Mehta2009}. For our system, this suppression factor of the three-body recombination is expected to be $1/2!$, given that only two of the three particles are BEC atoms~\cite{Mikkelsen2015}. We have included this prefactor in Eq.~\eqref{eq:L_3}.

As the scattering length is increased, the three-body loss rate is expected to saturate~\cite{Incao2004,Rem2013}. This occurs when the scattering length becomes comparable to the interparticle distance, with the latter becoming the only physical length scale of the system. To account for this saturation, the three-body loss rate is modelled as 
\begin{equation}
    L_{3} = \left(\frac{1}{L_{\text{max}}} + 2!\frac{m}{n_lC\hslash a^4}\frac{\sin(s_0\ln(a/a_-))^2 + \sinh(\eta_*)^2}{\sinh(2\eta_*)}\right)^{-1},
    \label{eq:L_eff}
\end{equation}
where $L_{\text{max}}$ is the saturation value. This effective three-body loss rate is fitted to the experimental data and plotted in Fig.~\ref{fig:K3_coeffs}. Importantly, this fit provides $L_3$ at any scattering length which can be used to determine the medium atom number with Eq.~\eqref{eq:Diff_final}.

The fit captures the data well, from the small scattering length behavior given by Eq.~\eqref{eq:L_3}, toward the saturation at large scattering lengths. The fitted value, $C = 4.8(1.2) \times 10^2$, is somewhat higher than the expected theoretical prediction of $\sim1.33\times 10^2$, for two identical bosons interacting with a third boson of equal masses~\cite{Helfrich2010}. However, the parameters for the Efimov resonance have an impact on the extracted $C$ coefficient and the fitted value of $C$ can vary significantly for different combinations of $\eta_*$ and $a_-$. We conclude that the systematic uncertainty of $C$ is much larger than the statistical uncertainty provided above. To reduce this uncertainty, further experimental and theoretical work is required to provide strict constraints on the values of $a_-$ and~$\eta_*$. 

The extracted experimental saturation value in Eq.~\eqref{eq:L_eff} for the three-body loss rate is $L_{\text{max}} = 3.06(0.63)\times 10^{-23}~\text{cm}^6/\text{s}$. In the case of a thermal system, this saturation is related to the temperature of the system and can be compared to the theoretical prediction of Ref.~\cite{Rem2013}. We substitute the thermal energy with the typical energy scale of the BEC $E_n$ in Ref.~\cite{Rem2013} and thus obtain
\begin{equation}
    L_\text{max} = \frac{36\sqrt{3}\pi^2\hslash^5}{E_n^2m^3},
    \label{eq:L_max}
\end{equation}
which provides a saturation value for our system analogous to Refs.~\cite{Smith2014,Makotyn2014,Chevy2016,Eigen2017}. This yields a saturation value of $2.22(0.31)\times 10^{-23}~\text{cm}^6/\text{s}$, matching the measured value within the statistical uncertainties.

Importantly, the experimental values of $C$ and $L_\text{max}$ take the density and the properties of the BEC into account. However, further effects due to the large density~\cite{wacker2018,Chapurin2019} such as secondary collisions may influence the result. In the following section, it is shown that only for values of the scattering length $|a|>10^3$ (corresponding to $|1/k_na|<0.9)$) additional collisional processes take place. The value of $C$ should therefore remain unaffected by these processes, while the value of $L_\text{max}$ may be influenced by them.

\section{Dominant loss mechanisms in atom number determination}
\label{sec:loss_meachnisms}

To gain a more detailed understanding of the loss processes at play, the coefficients $\eta_3$ and $\eta_2$, defined in Sec.~\ref{sec:reconstruction}, should be examined in more detail. These loss coefficients are central for reconstructing the medium atom number from Eq.~\eqref{eq:N_final} and Eq.~\eqref{eq:true_atomnumber_ejec} and for evaluating the atom number through the differential equation in Eq.~\eqref{eq:Diff_final}. The loss coefficients are expected to be $\eta_3 \approx 2$ and $\eta_2 \approx 1$, corresponding to the number of medium atoms lost per impurity atom due to three- and two-body losses, respectively. In this section, we present our experimental results for these coefficients.

A total loss coefficient, including both loss processes occurring during the entire sequence, can be defined as

\begin{figure}[b!]
    \centering
    \includegraphics[width=\columnwidth]{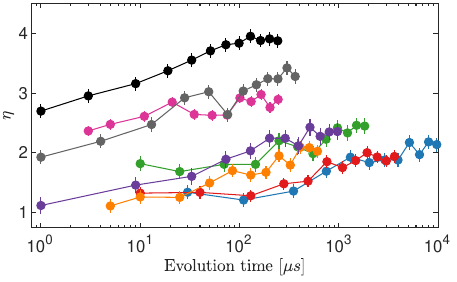}
    \caption{Extracted total medium atom loss coefficient, $\eta$, as a function of evolution time for different inverse interaction strengths, $1/k_na = -4.2$ (blue), $-3.2$ (red), $-2.4$ (green), $-1.6$ (orange), $-0.9$ (purple), $-0.4$ (magenta), $-0.2$ (gray) and $-0.1$ (black). For weak interaction strengths, we observe the expected loss of 1 and 2 medium atoms per impurity due to two-body collisions at short and three-body collisions at long times, respectively. For large interaction strengths, the losses are significantly increased, even at short times.}
    \label{fig:Loss_coeff}
\end{figure}

\begin{equation}
    \eta = \frac{N_0(1-\chi)-N_\text{obs}(t)}{N_0\chi}.
    \label{eq:Loss_coeff_general}
\end{equation} 
This corresponds to the difference in the initial and observed medium atom number divided by the initial number of impurities in the system. Thus, this general loss coefficient corresponds to the loss coefficients $\eta_2$ and $\eta_3$ in the limits of short and long evolution times, respectively. Experimentally, these two loss coefficients are evaluated at very short and long evolution times from the observed medium atom number as discussed in Sec.~\ref{sec:reconstruction} based on Fig.~\ref{fig:decay_curve}. However, for intermediate evolution times, the total loss coefficient contains contributions from both loss processes.

The total loss coefficient is shown in Fig.~\ref{fig:Loss_coeff} as a function of evolution time for different impurity-medium interaction strengths. For weak interaction strengths, the loss coefficient is indeed observed to be close to 1 for short evolution times and increases to 2 for long evolution times, as expected. However, for strong interactions, the losses are significantly increased, and the average number of medium atoms lost per impurity for long times increases to values between 3 and 4. In this regime, even the losses for very short evolution times are significantly increased.

The observed increase of lost medium atoms at large interaction strengths in Fig.~\ref{fig:Loss_coeff}, does not coincide with a sudden increase of the three-body loss rate coefficient in Fig.~\ref{fig:K3_coeffs} which would indicate a loss beyond the universal $a^4$ dependence affiliated with higher order loss processes. This is not surprising, since the scattering threshold for the impurity-medium-medium Efimov trimer is expected to be on the order of $|a_-| =3\times 10^5\ a_0$, well above the values accessible here~\cite{jorgensen2016,Levinsen2015}. Instead, we attribute the additional observed loss to secondary collisions in the system~\cite{zaccanti2009,Schuster2001,braaten2013}.

The loss coefficient, $\eta_3$, can also be obtained from the off-resonant atom number in ejection spectroscopy (see Sec.~\ref{sec:spectroscopy}). This corresponds to using $N_\text{obs} = N_\text{off-res}$ in Eq.~\eqref{eq:Loss_coeff_general}, leading to

\begin{figure}[t!]
    \centering
    \includegraphics[width=\columnwidth]{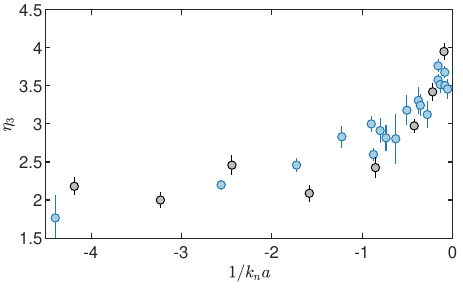}
    \caption{Extracted $\eta_3$ loss factor as a function of the inverse interaction strength, $1/k_na$, calculated from the observed atom number off-resonance for ejection spectra (blue points). The final extracted general loss factor for each curve in Fig.~\ref{fig:Loss_coeff} is also plotted (gray points).}
    \label{fig:C0}
\end{figure}
\begin{equation}
    \eta_3 = \frac{N_0(1-\impfrac)-N_{\text{off-res}}}{N_0\impfrac}.
    \label{eq:loss_eta3}
\end{equation}
The obtained $\eta_3$ loss coefficients from Fig.~\ref{fig:Loss_coeff}~\footnote{Here $\eta_3$ corresponds to the value of $\eta$ at the longest available evolution time for each interaction strength.} and Eq.~\eqref{eq:loss_eta3} are shown in Fig.~\ref{fig:C0} as a function of the inverse impurity-medium interaction strength. At weak interaction strengths, the extracted loss coefficient fits with the expected $\eta_3 = 2$ behavior but starts to strongly increase when approaching unitarity. Importantly, the obtained $\eta_3$ loss coefficients from the ejection spectra are observed to be in excellent agreement with those from Fig.~\ref{fig:Loss_coeff}.

\section{Conclusion} 
\label{sec:conclusion}

In summary, we have investigated three-body loss processes of impurities embedded in a Bose-Einstein Condensate close to a Feshbach resonance by using an even faster loss process to probe the system. This involves the ejection of the impurity atoms to a state that decays due to two-body collisions with the BEC. It constitutes a new tool in loss spectroscopy which allows for the investigation of system properties linked to intrinsic resonant loss processes. 

It was shown that the medium atom number in the BEC can be reconstructed based on the outcome of the ejection measurement, both for time-dependent loss measurements and spectroscopic ejection measurements. This medium atom number gives access to important system parameters such as the density of the sample and its characteristic energy.  

Both the loss and the spectroscopic method allow for a calculation of the atomic density, and we provide a detailed comparison as a function of interaction strength and evolution time. We discuss discrepancies at large interaction strengths, highlighting the limitations of modelling the three-body processes in this regime. Here, the spectroscopic measurements provide the most direct experimental access to the atom number and density, and thus provide a benchmark.

A theoretical model of the three-body loss process in the impurity limit was applied to the experimental data, providing measurement of the three-body loss rate coefficient $L_3$ at strong impurity-medium interactions. The coefficient was investigated as a function of the scattering length and observed to agree well with the expected universal behavior depending on $a^4$, and its expected saturation value. The three-body loss rate coefficient also allows for calculating the medium atom number of the system for any interaction strength and evolution time.

Finally, our method allows for an evaluation of the number of medium atoms lost per impurity atom as a function of the interaction strength and evolution time. It is found to exceed 2 at strong interactions, and an increase towards maximal values of 4 is observed when approaching large interaction strengths at $1/k_na = 0$. This increase was attributed to secondary collisions after the initial loss process.

The presented methods offer a new tool for investigating collisional processes in strongly interacting ultracold gases. Importantly, the impurity limit of mixtures is specifically exploited to isolate the loss channels. The loss process of two identical bosons interacting with a third boson investigated here shows that the extracted three-body loss rate coefficient is lower compared to investigations of single-component thermal gases of $^{39}$K~\cite{zaccanti2009,wacker2018}, as expected~\cite{Helfrich2010}. In future work it will be interesting to explore whether further effects may have to be considered when utilizing a BEC, such as the suppression of the three-body recombination at high densities~\cite{Chapurin2019}.

Our results may find use in experiments that use loss spectroscopy of $^{39}$K~\cite{jorgensen2016,hu2016,etrych2024}, but can also be extended to investigate solitons and quantum droplets~\cite{Lepoutre2016,semeghini2018,Skov2021}, Efimov physics~\cite{zaccanti2009,wacker2018,etrych2023}, and collisional avalanches in BECs~\cite{Schuster2001,braaten2013,hu2014}.

\section{Acknowledgements} 
The authors thank Adam Simon Chatterley for valuable discussions. We acknowledge support from the Danish National Research Foundation through the Center of Excellence “CCQ” (DNRF152) and by the Novo Nordisk Foundation NERD grant (Grantno. NNF22OC0075986).
\bibliography{analysis}

\end{document}